\documentclass[nofootinbib]{revtex4}
\usepackage{graphicx}
\usepackage{fancyhdr}
\usepackage{amsmath,amssymb}
\usepackage{color}

\begin{document}

\title{Direct searches of extra Higgs boson at future
colliders\footnote{%
The talk is based on Ref.~\cite{Kanemura:2015nza} in collaboration with
 S.~Kanemura and Y.J.~Zheng.
}}

%

\author{Hiroshi Yokoya\footnote{%
Electric address: {\tt hyokoya@gmail.com}; present address: Theory
Center, KEK, Tsukuba 305-0801, Japan}} 
\affiliation{Department of Physics, University of Toyama, Toyama
930-8555, JAPAN}

\begin{abstract}
We study direct searches of additional Higgs bosons in multi-top-quarks
 events at the LHC with the collision energy of 14~TeV as well as the
 International Linear Collider (ILC) with the collision energy of 1~TeV.
As a benchmark model, we consider two Higgs doublet models with a
 softly-broken discrete $Z_2$ symmetry, where the $t\bar t$ decay mode
 of additional  neutral Higgs bosons can be dominant if their masses are
 heavy enough. 
Thus, the multi-top-quarks events become an important probe
 of the extended Higgs sector at future colliders.
We estimate the discovery reach at the LHC and the ILC, and find that
the search at the ILC can survey the parameter regions where the LHC
 cannot cover.
\end{abstract}

\maketitle

\thispagestyle{fancy}


\section{Introduction}

A Higgs boson was discovered at the LHC.
Further measurements at the LHC have revealed its properties, such as its
mass, spin-parity, and couplings to the standard model (SM) particles. 
At the present experimental precision, it seems quite consistent with
the Higgs boson in the SM.
The determination of the whole structure of the Higgs sector is an
important task at future experiments, since various kinds of the Higgs
sector are predicted in models for the physics beyond the SM (BSM). 
To explore the whole structure of the Higgs sector at collider
experiments, two major approaches have been considered; the direct
searches of second, third and even more Higgs bosons, and the indirect
searches of the BSM effects through the deviations in the observed Higgs
boson properties by precision measurements. 

In general, it is argued that the direct searches are performed at the 
energy frontier experiments, such as the LHC, and the indirect searches
are performed at the precision frontier experiments, such as the
International Linear Collider~(ILC)~\cite{Behnke:2013lya}. 
In this talk, we discuss the direct searches of additional Higgs bosons
at the ILC~\cite{Kanemura:2015nza,Kanemura:2014dea} in addition to the
LHC, and their complementarity.
We consider the two Higgs doublet model (2HDM) as a benchmark model of
the extended Higgs sector.
Especially, we focus on the heavy (neutral) Higgs bosons which
predominantly decay into a $t\bar t$ pair resulting multi-top-quarks
events as a signal of such Higgs bosons.
We study the discovery potential of such events at the LHC and the ILC,
and compare the parameter regions to be explored by the searches at
these experiments.
We find that the parameter regions to be explored at the LHC and the ILC
are different, therefore, searches at the two experiments are
complementary to survey the wider parameter regions in the extended
Higgs sector. 

\section{Two Higgs Doublet Model}

We consider the 2HDM as a benchmark model of the extended Higgs sector.
The model has various phenomenologically plausible features, such that
the electroweak rho parameter is preserved to be unity at the
tree-level; sources of additional $CP$ phases are provided; and
additional four Higgs bosons are predicted.
Under the assumption of the $CP$ invariance in the Higgs sector for
simplicity, these can be identified as $CP$-even $H$, $CP$-odd $A$, and
a pair of charged $H^\pm$, in addition to the lighter $CP$-even $h$
which can be identified as the Higgs boson discovered at the LHC with
the mass of $m_h\simeq125$~GeV.
Flavour changing neutral currents whose existence is experimentally
severely constrained can be avoided at the tree level by imposing the
softly-broken $Z_2$ symmetry.

Yukawa interactions of additional Higgs bosons to the SM fermion are
given as 
\begin{align}
 -{\cal L}_{\rm Yukawa} &= \sum_{f=u,d,\ell}\left[
\frac{m_f}{v}\xi_h^f\bar{f}fh + \frac{m_f}{v}\xi_H^f\bar{f}fH
-i\frac{m_f}{v}\xi_A^f\bar{f}\gamma_5fA \right]\nonumber\\
& + \left\{
\frac{\sqrt{2}V_{ud}}{v}\bar{u}\left[m_u\xi_A^u{\rm P}_L
+m_d\xi_A^d{\rm P}_R\right]dH^+ +
 \frac{\sqrt{2}m_\ell}{v}\xi_A^\ell\overline{\nu_L}\ell_RH^+ + {\rm H.c.}
\right\},
\end{align}
where the scaling factor $\xi_\phi^f$ with $\phi=h,H,A$ and $f=u,d,\ell$
is a function of the mixing angles, $\alpha$ and $\beta$, in the mass
matrix of the $CP$-even and odd components of the Higgs doublets,
respectively.
In the 2HDM with the softly-broken $Z_2$ symmetry, there are four kinds
of the model for the Yukawa interactions, depending on the assignment of
the fermion parity under the $Z_2$ symmetry, namely, Type-I, Type-II,
Type-X and Type-Y.
Detail expressions for the scaling factors in each type of Yukawa
interactions can be found, e.g., in Ref.~\cite{Aoki:2009ha}.
For any type of Yukawa interactions, couplings of the additional Higgs
boson to top quarks are enhanced for small $\tan\beta$.
Thus, the signatures for such parameter regions can be heavy Higgs
bosons which dominantly decay into top quarks.
Therefore, the searches for the multi-top-quarks events can be a new
signals of additional Higgs bosons in future collider experiments.

\section{Multi-top quarks production at the LHC}

First, we study multi-top-quarks production at the LHC through the
production of additional Higgs boson(s).
The process with largest cross section is expected to be
\begin{align}
 pp\to t\bar t H(t\bar t A)\to t\bar t t\bar t, \label{eq:tth}
\end{align}
since it occurs via the strong interaction.

The other process can be pair production of $H$ and $A$, 
\begin{align}
 pp\to HA\to t\bar t t\bar t,\label{eq:ha}
\end{align}
followed by the decays of $H$ and $A$ into top quarks, 
and also the associated production of $H^\pm$ and $H$(A), which
subsequently decay into $t\bar b$($\bar t b$) and $t\bar t$
\begin{align}
 pp\to H^\pm H (H^\pm A) \to t\bar b t\bar t /\bar t b t\bar t,
 \label{eq:hhp}
\end{align}
resulting three top-quarks plus one bottom-quark final-states.

The hadronic cross sections for the above processes can be predicted in
perturbative QCD, and by multiplying the branching ratios of additional
Higgs bosons into top quarks, expected number of events can be
estimated~\cite{Kanemura:2015nza}. 
The cross sections of the production of a pair of additional Higgs
bosons, (\ref{eq:ha}) and (\ref{eq:hhp}), do not depend on $\tan\beta$,
but that of the process (\ref{eq:tth}) does through the top Yukawa
coupling. 
The largest cross sections are realized for the mass of additional
Higgs bosons at around 350~GeV. 
Below this value, the branching ratio into $t\bar t$ is suppressed
because one of the top quarks is forced to be off-shell, while above
that value the production cross sections of additional Higgs bosons
decrease. 
For $\tan\beta=1$, the cross section of the four top-quarks production
can be at most 6~fb (50~fb) for the LHC 8~TeV (14~TeV).

The CMS Collaboration has set the limit on the cross section at 8~TeV as
$\sigma_{4t}(8~{\rm TeV})\le32$~fb at the 95\% CL (confidence
level)~\cite{Khachatryan:2014sca}, by observing the 
lepton plus multi-jets events.
On the other hand, the SM prediction to the four top-quarks production
at the LHC 8~TeV is about 1~fb.
Therefore, the limit is not enough to constrain the parameter regions in
the 2HDM. 

We study the prospect of measuring four top-quarks production by the
production of additional Higgs bosons at the LHC with the collision
energy of 14~TeV in the 2HDM. 
The total cross section within the SM is estimated to be $\sigma_{\rm
SM}=15$~fb with $\delta\sigma_{\rm SM}=4$~fb~\cite{Bevilacqua:2012em}. 
We follow the signal-to-background analysis in Ref.~\cite{Lillie:2007hd}.
In their analysis, the background rate of $B=7.2$~fb is obtained with
the signal efficiency of $\epsilon=0.03$.
By taking into account the statistical and systematical
uncertainties for the signal, SM and background processes, the
accuracy of measuring the signal cross section $\sigma_S$ can be
estimated to be
\begin{align}
 \frac{\delta \sigma_S}{\sigma_S} = 
\sqrt{\frac{\left(\sigma_{S}+\sigma_{\rm
 SM}\right)\epsilon+B}{\sigma^2_{S}\epsilon^2{\cal L}}
+\frac{\delta\sigma_{\rm
 SM}^2\epsilon^2+(\delta{B})^2}{\sigma_{S}^2\epsilon^2}},
\label{eq:acc}
\end{align}
where $\delta{B}$ denotes the systematic uncertainty of the background
rate which is taken to be $\delta{B}=0.05B$ in our analysis.
By solving Eq.~(\ref{eq:acc}), we find that the signal total cross
section $\sigma_S$ greater than 25~fb (63~fb) is required to achieve
$\delta\sigma_S/\sigma_S<0.5$ (0.2) with the integrated luminosity of
${\mathcal L}=300$~fb$^{-1}$.
In our setup, the total uncertainty is dominated by the systematic
uncertainty from the background.

\begin{figure}[t]
 \begin{center}
  \includegraphics[width=.35\textwidth,clip]{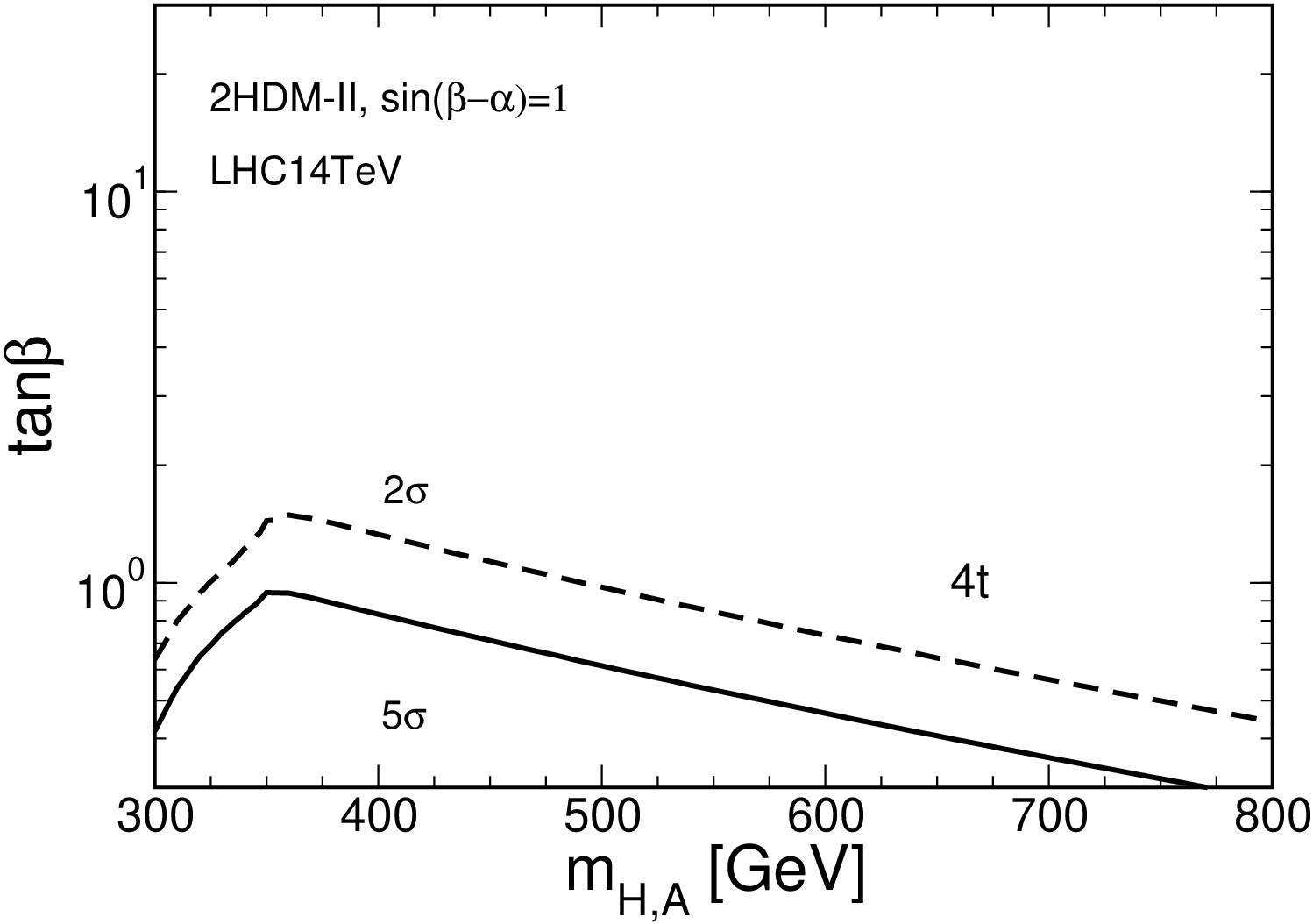}
  \includegraphics[width=.35\textwidth,clip]{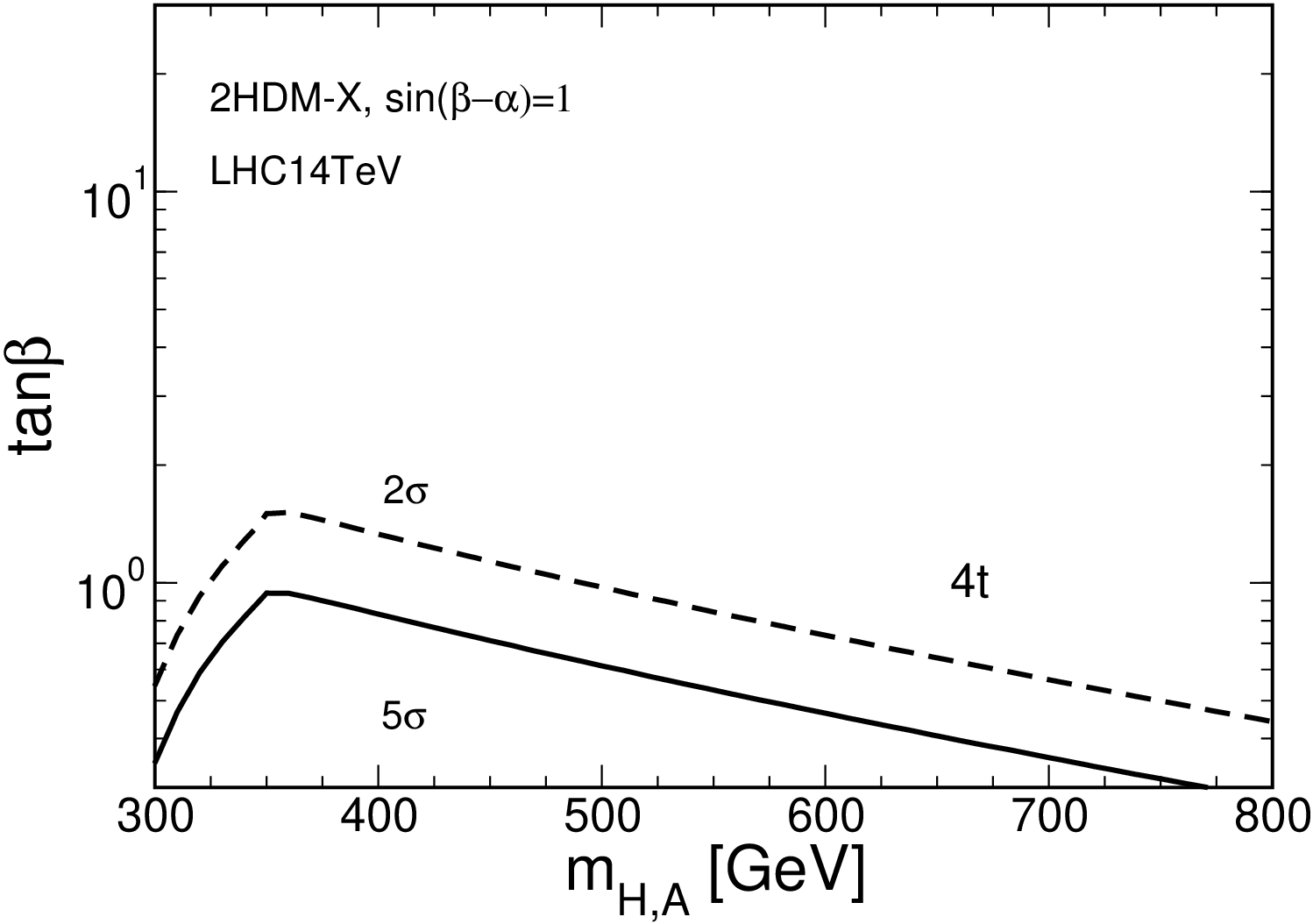}
 \end{center}
 \caption{Contour plot for the discovery reach of the four top-quarks
 events at the LHC with 3000~fb$^{-1}$ data in the 2HDM Type-II and
 Type-X. 
Discovery regions at the $2\sigma$ [$5\sigma$] CL are shown in the
 dashed [solid] lines.}\label{fig:LHC4top}
\end{figure}

In Fig.~\ref{fig:LHC4top}, we show the parameter regions in which the
additional Higgs bosons contribution can be detected in the four
top-quarks events at the $2\sigma$ and $5\sigma$ CL in the Type-II and
Type-X 2HDM.
The dependence on the type of Yukawa interactions is small.
We find that only $\tan\beta\lesssim1.5$ regions can be probed at the
$2\sigma$ CL at most. 
Since these regions are constrained by flavor experiments, the LHC
searches in the four top-quarks events may not have significant impact
on exploring the parameter regions in the 2HDMs.

\section{Multi-top quarks production at the ILC}

Second, we consider the multi-top-quarks production at the ILC.
In the 2HDM, the four top-quarks final-state is generated via the pair
production of $H$ and $A$, 
\begin{align}
 e^+e^-\to HA,
\end{align}
which is kinematically accessible for $\sqrt{s}\ge m_H + m_A$, 
and via the single production of $H$ or $A$~\cite{Kiyoura:2003tg}, 
\begin{align}
 e^+e^-\to t\bar tH (t\bar tA),\label{eq:ttha}
\end{align}
which is allowed even in the case with $\sqrt{s}\le m_H + m_A$. 
The $HA$ pair production cross section does not depend on $\tan\beta$ at
the tree level, and thus the four top-quarks production rate depends on
$\tan\beta$ only through the decay branching ratio of $H$ and $A$. 
On the other hand, the single production process can be increased by the
enhanced Yukawa coupling of $H$ and $A$ to top
quarks~\cite{Kanemura:2014dea}.

We perform a simulation analysis on four top-quarks production at
the ILC.
The SM backgrounds processes, such as $e^+e^-\to t\bar t$, $t\bar t
b\bar b$, $t\bar t \ell^+\ell^-$ are taken into account.
The SM four top-quarks production cross section is negligibly small.
Our simulation analysis includes the decay of top quarks, QCD showering,
hadronization, jet clustering, flavor tagging, detector acceptance and
momentum resolution effects, etc~\cite{Kanemura:2015nza}. 
We make use of the designed detector performance at the ILC
experiment~\cite{Behnke:2013lya}.

Through the decay of top quarks, the four top-quarks events can be
observed in all-hadronic, single lepton plus jets plus missing momentum,
dilepton plus jets plus missing, trilepton plus jets plus missing,
tetralepton plus jets plus missing channels. 
In principle, optimal selection cuts may be constructed for each channel.
However, in this work, we propose an inclusive analysis which take into
account all the channels at once for simplicity.
To extract the signal events out of the SM background, we impose following
selection cuts; (i) small thrust, $T<0.77$, (ii) $N_{bj}\ge3$,
and (iii) large multiplicity of hard objects, $N_T\equiv2N_{\rm
lep}+N_{\rm jet}\ge10$, where $N_{\rm lep}$ is the sum of the number of
isolated leptons ($e$ and $\mu$) and the number of hadronic $\tau$-jets,
$N_{bj}$ ($N_{j}$) is a number of $b$-tagged (light-flavor) jets in a
event.

The event distributions of the signal and background processes in the
kinematical variables used for the selection cuts as well as the
background reduction and signal detection efficiencies can be found in
Ref.~\cite{Kanemura:2015nza}.
We obtain that the signal events can be extracted with the detection
efficiency of 40-50\% depending on the mass of additional Higgs bosons,
while only about 50~ab of the background events is remained.
The accessible value of the total cross section is evaluated to be 
0.034-0.048~fb (0.11-0.15~fb) at the $2\sigma$ ($5\sigma$) CL, depending
on the mass of additional Higgs bosons.

\begin{figure}[t]
 \begin{center}
  \includegraphics[width=.35\textwidth,clip]{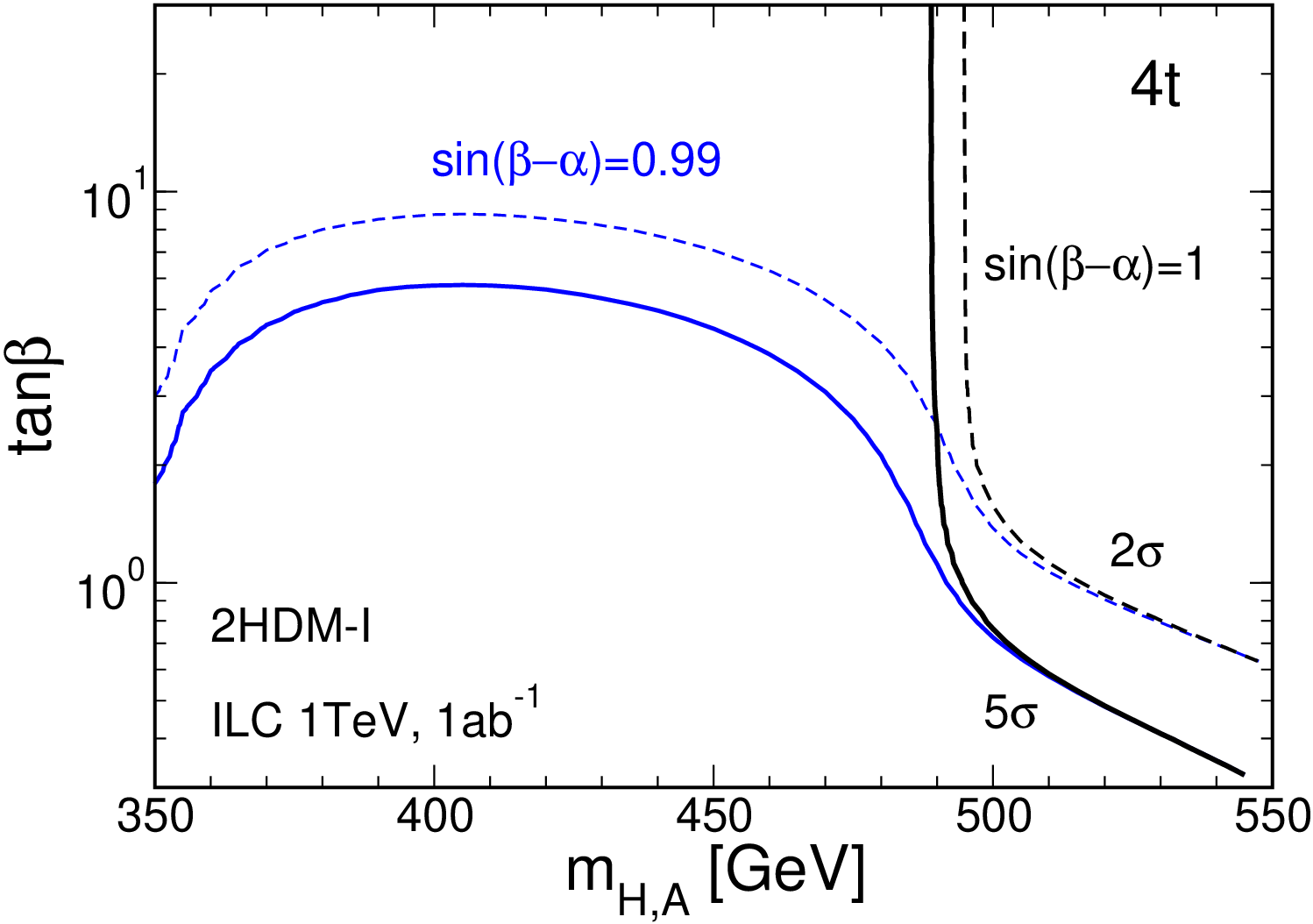}
  \includegraphics[width=.35\textwidth,clip]{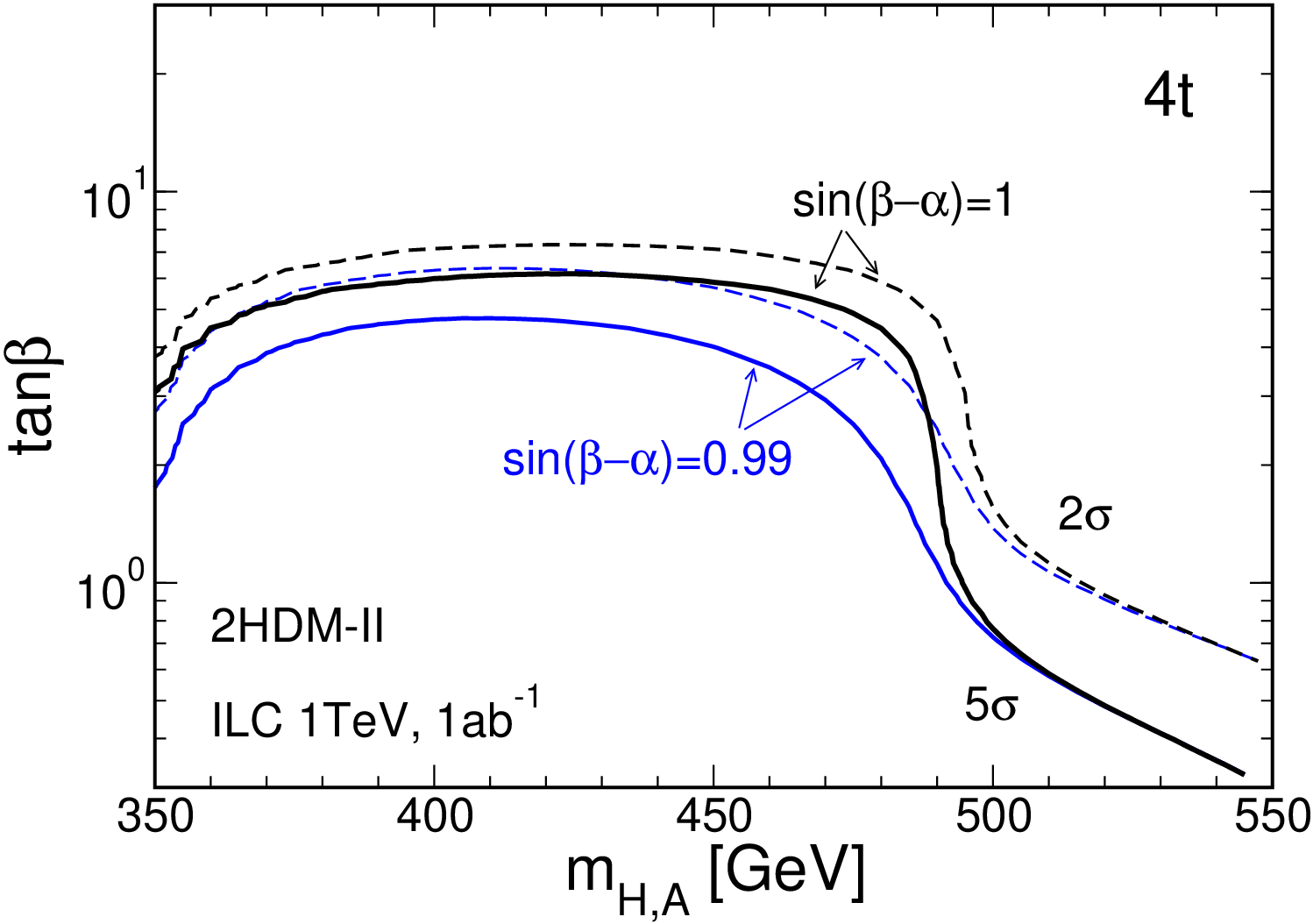}
 \end{center}
 \caption{Contour plot for the discovery reach of the four top-quarks
 events at the ILC 1~TeV with 1~ab$^{-1}$ data in the 2HDM Type-I and
 Type-II. 
 Discovery regions at the $2\sigma$ [$5\sigma$] CL are shown in dashed
 [solid] lines.}\label{fig:ILC4top}
\end{figure}

We evaluate the parameter regions in the 2HDMs where the four top-quarks
events can be detected at the ILC 1~TeV run with the integrated
luminosity of 1~ab$^{-1}$.
We take into account the statistical uncertainties only, since the
systematical ones can be expected to be well under control at lepton
colliders. 
In Fig.~\ref{fig:ILC4top}, we plot the solid-line (dashed-line) contours 
in the $(m_{H,A}, \tan\beta)$ plane where the four top-quarks events can
be detected at the $2\sigma$ ($5\sigma$) CL at the ILC 1~TeV with ${\cal
L}=1$~ab$^{-1}$. 
Results for Type-I (left) and Type-II (right) 2HDMs are shown for
example.
For Type-II, $\tan\beta$ up to about 7 (6) can be probed at the $2\sigma$
($5\sigma$) CL for $m_{H,A}\lesssim500$~GeV. 
For Type-I, the events can be detected for any value of $\tan\beta$,
as long as $m_{H,A}\lesssim500$~GeV.

The result for Type-I strongly depends on our choice of taking
$\sin(\beta-\alpha)=1$ where no $H\to WW$, $ZZ$, $hh$ or $A\to Zh$
decay is induced.
If we take $\sin(\beta-\alpha)=0.99$, these decay modes become non-zero,
and even dominant for larger $\tan\beta$~\cite{Kanemura:2014bqa}. 
In this case, the four top-quarks events can be observed up to about
$\tan\beta\simeq 8$~(5) at the $2\sigma$ ($5\sigma$) CL.
The discovery reaches for $\sin(\beta-\alpha)=0.99$ are also shown in
blue lines in Fig.~\ref{fig:ILC4top}.

The searches for multi-top-quarks events at the LHC and the ILC 
explore the different parameter regions in the 2HDM.
At the LHC Run-II and 3000~fb$^{-1}$, only the parameter regions with
$\tan\beta\lesssim1.5$ can be surveyed.
On the other hand, at the ILC 1~TeV, we find that the parameter regions
with larger $\tan\beta$ can be surveyed as long as
$m_{H,A}\lesssim500$~GeV. 
For Type-II, the parameter regions with $\tan\beta\simeq 7$ can be
explored.
For Type-I, because the decay mode into a top-quark pair is dominant for
any value of $\tan\beta$, the detection of the four top-quarks events is
anticipated in any value of $\tan\beta$. 
This result is modified to $\tan\beta\lesssim9$ in the case of
$\sin(\beta-\alpha)=0.99$.

\section{Summary}

We have studied the direct searches of the additional Higgs bosons in
multi-top-quarks events at the LHC and the ILC.
As a benchmark model of the extended Higgs sector, we have considered
 the two Higgs doublet models with softly-broken discrete symmetry. 
At the LHC, the systematical uncertainties of estimating the SM
backgrounds dominate the uncertainty, and therefore, the searches can
only survey the parameter regions with $\tan\beta\lesssim1.5$ which is
disfavored by experimental constraints.
At the ILC, we have shown that the parameter regions with
$\tan\beta\lesssim8$-15 can be surveyed, depending on the type of Yukawa
interactions, although the mass reach is almost limited by the collision
energy. 
The direct searches for the mulit-top-quark events at the LHC and the
ILC can be complementary to explore the wider parameter regions in the
2HDMs.

\bigskip 
\begin{acknowledgments}
The author would like to thank Shinya Kanemura and Ya-Juan Zheng for 
fruitful collaborations.
\end{acknowledgments}

\bigskip 

\end{document}